# Effects of normal and oblique incidence on zero-$\bar{n}$ gap in periodic lossy multilayer containing double-negative materials

Alireza Aghajamali and Mahmood Barati [*]

Department of Physics, Science and Research Branch, Islamic Azad University, Fars, Iran

**Abstract**

In this paper the transmission of electromagnetic waves through a one-dimensional lossy photonic crystal consisting of layers with negative and positive refractive indices is investigated. The behavior and characteristics of the bandwidth, the depth and the central frequency of the zero-$\bar{n}$ gap for different incidence angles, polarization, loss factor and ratios of thickness of the layers are studied. The results show that the gap is very sensitive to the incidence angle, polarization and the thickness ratio, but it is nearly insensitive to small loss factor. Such properties are quite useful in designing new types of edge filters and other optical devices in microwave engineering.

## 1. Introduction

Photonic crystals (PCs) are artificial dielectric or metallic structures in which the refractive index changes periodically. PCs have attracted much interests due to their novel electromagnetic wave characteristics and important scientific and engineering applications and have received attentions of many researchers in recent decades. Recently, with possibility of producing metamaterials, PCs with metamaterials (photonic crystals with positive- and negative refractive indices) are made. The properties of the conventional PCs band gap called "Bragg-gap" have been extensively studied by many authors [1-5]. It has been shown that the gap strongly depends on the angle of incidence, lattice constant, polarization and disorder. The properties of PCs with double-negative (DNG) materials have also been investigated in the last few years [6-13].

In an interesting report by Li et al. [7] the optical properties of a one-dimensional (1D) PC with DNG and double-positive (DPS) layers are investigated. They showed that such materials indicate an additional gap, called the "zero-$\bar{n}$ gap", with properties quite different from the ordinary Bragg gap. The spectral position and the width of the new gap are invariant upon a change of scaling and are also independent of the incidence angle and polarization [7,8]. In a detail study of the zero-$\bar{n}$ gap Daninthe et al. [9] and Awasthi et al. [12] showed that only for the s-polarized wave such structure indicates an efficient omnidirectional gap.

In this paper the transmission of electromagnetic waves through a 1D lossy PC consisting of layers with negative- and positive-index material is studied and the effects of the loss factor, the angle of incidence, polarization and the layer thickness ratio on the characteristics of the zero-$\bar{n}$ gap are investigated. The paper is organized as follows: the PC structure and the theoretical formulation (characteristic matrix method) are described in Section 2, the numerical results and discussions are presented in Section 3 and the paper is concluded in Section 4.

---
[*] Corresponding author. Tel: +98 711 6249560 ; fax: +98 711 2280926.
 *Email address:* barati@susc.ac.ir (M. Barati)

## 2. Structure design and characteristic matrix method

The 1D PC under study, consisting of layers of DNG (layer A) and DPS (layer B) materials, located in the free space is shown in Fig. 1, where DNG material is dispersive and dissipative.

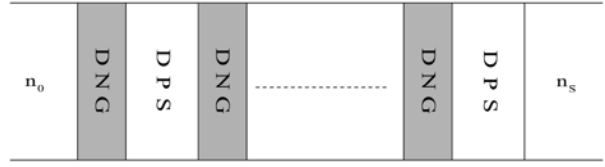

Figure 1. Schematic of 1D PC structure, consisting periodic layers of DNG-DPS materials.

Our calculations are based on the characteristic matrix method [14,15], which is the most effective to analyze the transmission properties of 1D PCs. The characteristic matrix $M[d]$ for TE waves at incidence angle $\theta_0$ from vacuum to a 1D PC structure is given by [14]:

$$M[d] = \prod_{i=1,2} \begin{bmatrix} \cos\gamma_i & \dfrac{-i}{p_i}\sin\gamma_i \\ -ip_i \sin\gamma_i & \cos\gamma_i \end{bmatrix} \quad (1)$$

where $\gamma_i = (\omega/c) n_i d_i \cos\theta_i$, $c$ is speed of light in vacuum, $\theta_i$ is the angle of refraction inside the layer with refractive index $n_i$ and $p_i = \sqrt{\varepsilon_i/\mu_i}\cos\theta_i$, where $\cos\theta_i = \sqrt{1-(n_0^2 \sin^2\theta_0/n_i^2)}$. The characteristic matrix for an $N$ period structure is therefore $[M(d)]^N$. The transmission coefficient of the multilayer is given by:

$$t = \frac{2p_0}{(m_{11} + m_{12} p_s) p_0 + (m_{21} + m_{22} p_s)}. \quad (2)$$

Here $m_{ij}\ (i,j=1,2)$ are the matrix elements of $[M(d)]^N$, where $p_0 = n_0 \cos\theta_i$ and $p_s = n_s \cos\theta_s$. The treatment for TM waves can be obtained by using these previous expressions with $p_i = \sqrt{\mu_i/\varepsilon_i}\cos\theta_i$, $p_0 = \cos\theta_i/n_0$ and $p_s = \cos\theta_s/n_s$.

The permittivity and permeability of the layer A with negative refracting index in the microwave region are complex and are given by [7],

$$\varepsilon_A = 1 + \frac{5^2}{0.9^2 - f^2 - if\gamma} + \frac{10^2}{11.5^2 - f^2 - if\gamma} \ ; \quad (3)$$

$$\mu_A = 1 + \frac{3^2}{0.902^2 - f^2 - if\gamma} \ ; \quad (4)$$

where $f$ and $\gamma$ are frequency and damping frequency, respectively, given in GHz. Plots of the real parts of the permittivity and permeability of layer A, $\varepsilon'_A$ and $\mu'_A$, versus frequency for $\gamma = 0$ GHz and large damping frequency ($\gamma = 1$ GHz) are shown in Fig. 2. As it is seen from the figure there exists three distinct regions. In the first region, $\varepsilon'_A$ and $\mu'_A$ are both negative (DNG material). In the second region, $\varepsilon'_A < 0$ but $\mu'_A > 0$ (single-negative (SNG) material extending to the epsilon-negative (ENG) material). In the third region, both $\varepsilon'_A$ and $\mu'_A$ are positive (DPS material). As it is seen, the real parts of $\varepsilon_A$ and $\mu_A$ are both negative for small and large loss factors for $f < 3$ GHz (the region of our interest).

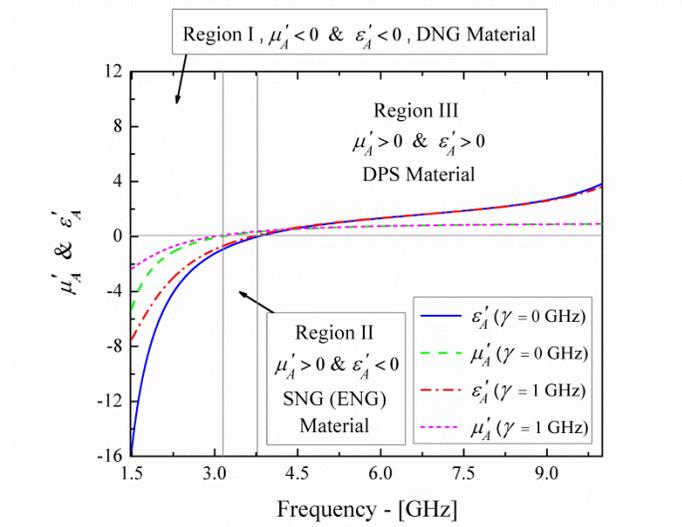

Figure 2. The real parts of permittivity and permeability of layer A, $\varepsilon'_A$ and $\mu'_A$, versus frequency for two different loss factors.

The second layer (DPS material) is assumed to be a vacuum layer with $\varepsilon_B = \mu_B = 1$. The behaviour of the width of the band gap as a function of the number of lattice periods, $N$, for different loss factors, incidence angles and polarizations are shown in Fig. 3. It is seen that the width of the band gap is almost independent of the number of periods for $N \geq 16$. Accordingly, the presented numerical calculations are performed for $N = 16$. It is also observed that the width of the band gap increases as the loss factor increases.

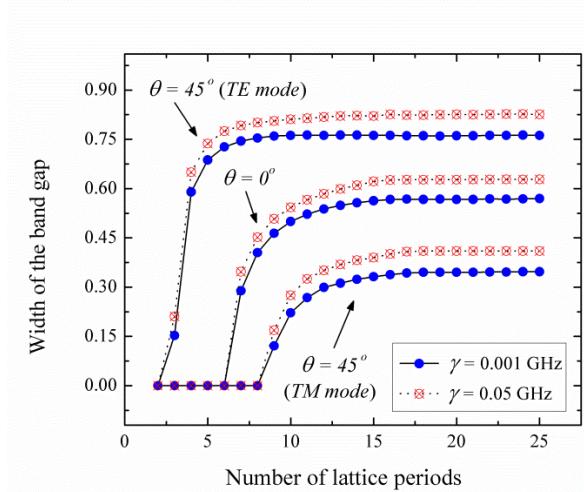

Figure 3. The width of the zero-$\bar{n}$ gap versus the number of the lattice periods for different loss factors, incidence angles and polarizations.

## 3. Results and discussion

In the first part, the transmission of electromagnetic waves through the 1D lossy PC for $d_A = 8$ mm and $d_B = 24$ mm was investigated. The effect of the loss factor on the zero-$\bar{n}$ gap for normal incidence angle is shown in Fig. 4. It is seen that as the loss factor increases the width of the band gap increases, the depth decreases, but the central frequency nearly remains unchanged. Moreover, the figure indicates that the transmittance is very sensitive to the loss factor and decreases when $\gamma$ increases. The dependence of the zero-$\bar{n}$ gap on the incidence angle, for small and large loss factors ($\gamma = 0.001$ GHz and $\gamma = 0.05$ GHz) for TE-polarized and TM-polarized waves are shown in Fig. 5 and Fig. 6, respectively. As it is seen from Fig. 5, the width, the depth of the gap and the central frequency ($f_c$) for TE waves increase as the incidence angle increases for both loss factors. Such a structure behaves as an omnidirectional gap and which is unaffected by the loss factor in the frequency range of 1.833–2.379 GHz. Such behavior has been reported by Awasthi et al. [12] in a similar non dispersive photonic crystal.

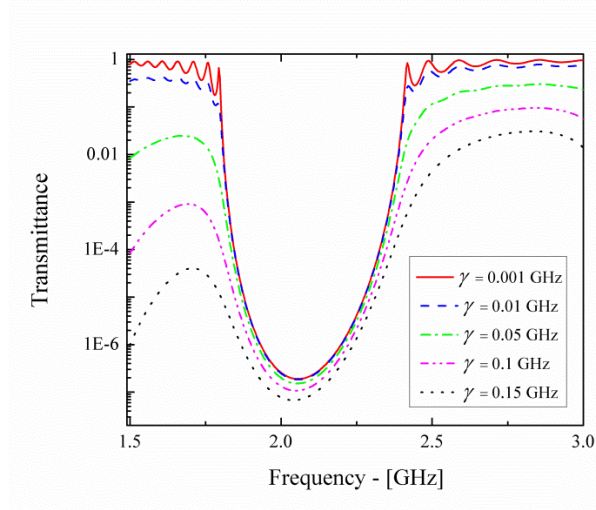

Figure 4. The transmission spectra for normal incidence angle and for different loss factors.

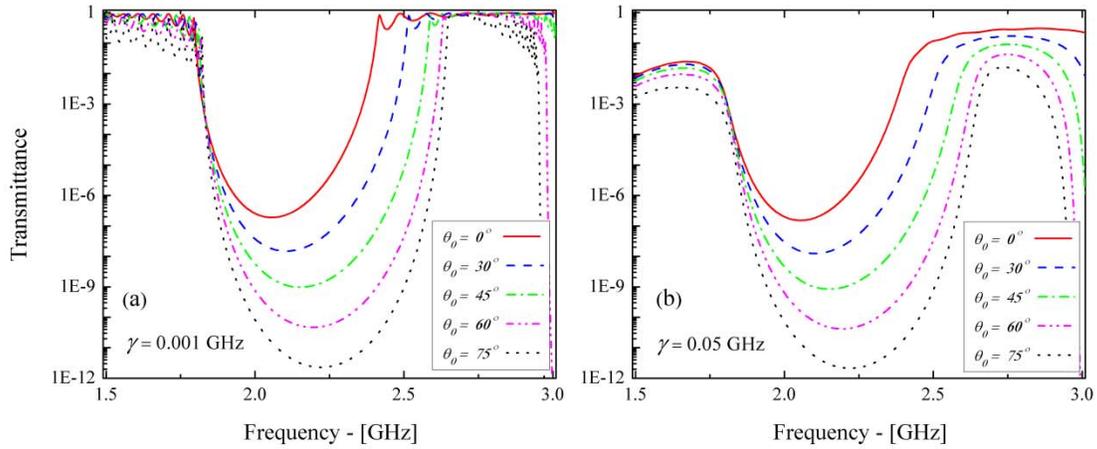

Figure 5. The transmission spectra of TE-polarized waves as a function of the incidence angle for (a) $\gamma = 0.001$ GHz (b) $\gamma = 0.05$ GHz.

The results for the TM-polarized wave are quite different, where the gap disappears for the incidence angle greater than $\approx 50°$. The width and the depth of the gap decreases and $f_c$ shifts toward the higher frequencies as the incidence angle increases for small and large loss factors as shown in Fig. 6.

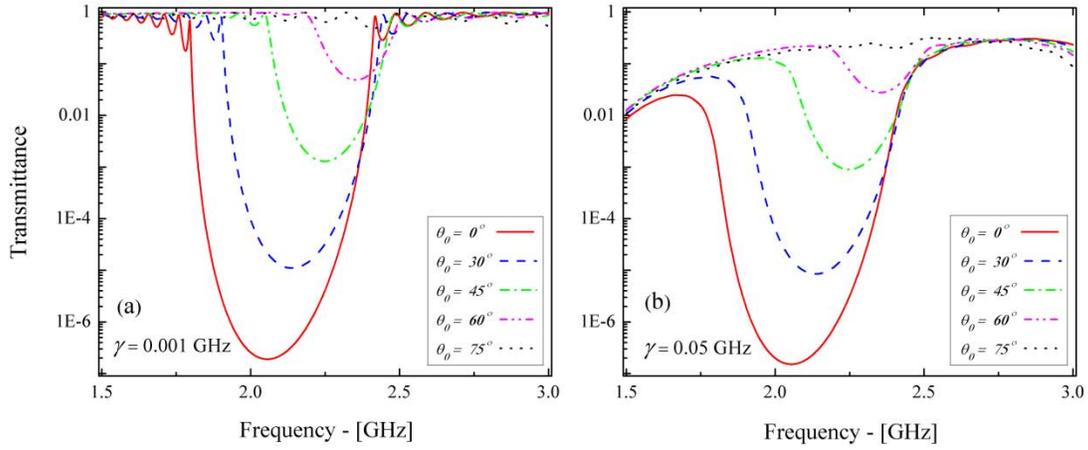

Figure 6. The transmission spectra of TM-polarized waves as a function of the incidence angle for (a) $\gamma = 0.001$ GHz (b) $\gamma = 0.05$ GHz.

In more details, the bandwidth of the gap and $f_c$ for both TE and TM waves as a function of incidence angle for two different loss factors have been presented in Fig. 7 and Fig. 8, respectively. The figures indicate that the width of the band gap is very sensitive to the loss factor but the central frequency is almost unaffected by $\gamma$ for both polarizations.

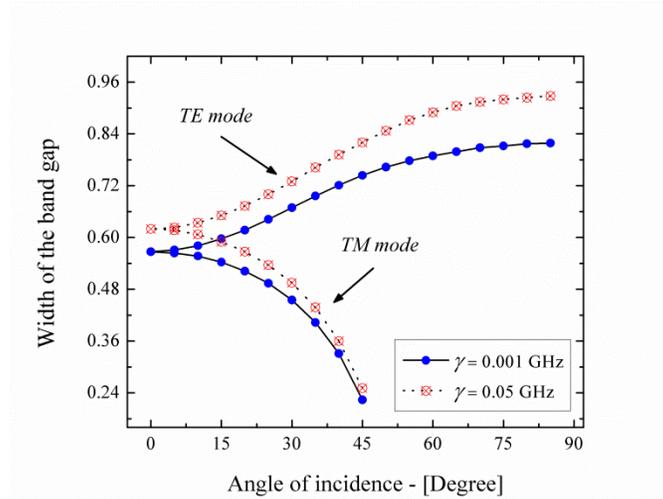

Figure 7. The width of the zero-$\bar{n}$ gap as a function of the angle of incidence for both polarizations and for two different loss factors.

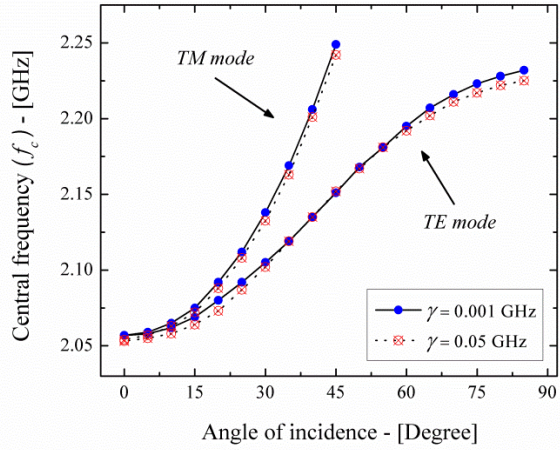

Figure 8. The central frequency of the zero-$\bar{n}$ gap as a function of the angle of incidence for both polarizations and for two different loss factors.

In the second part, the effect of the ratio of the thickness of layers on the zero-$\bar{n}$ gap is studied. The width, the depth and the central frequency of the zero-$\bar{n}$ gap as a function of the thickness ratio of the layers ($d_B/d_A$) for the normal incidence and for $d_A = 8$ mm was studied. As shown in Fig. 9 the central frequency of the zero-$\bar{n}$ gap decreases as the thickness ratio increases. The numerical results for the width and the depth of the zero-$\bar{n}$ gap as a function of the thickness ratio for small and large loss factors are shown in Fig. 10 and Fig. 11, respectively.

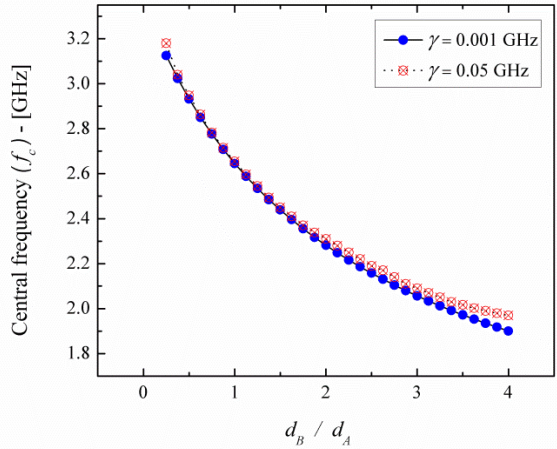

Figure 9. The central frequency of the zero-$\bar{n}$ gap versus the thickness ratio for two different loss factors.

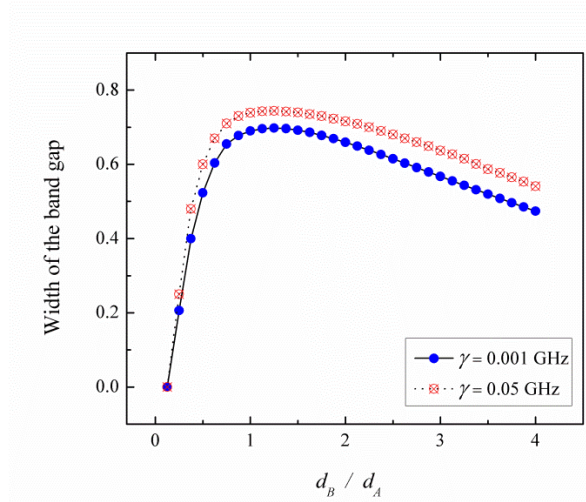

Figure 10. The width of the zero-$\bar{n}$ gap versus the thickness ratio for two different loss factors.

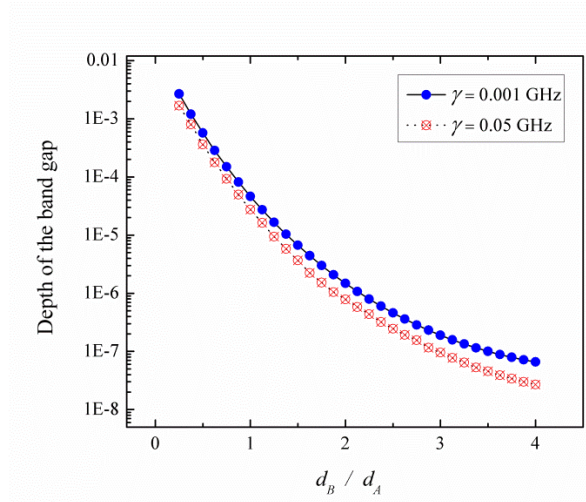

Figure 11. The depth of the zero-$\bar{n}$ gap on the thickness ratio for two different loss factors.

As it is seen from Fig. 10, the width nonlinearly depends on the thickness ratio and indicates a maximum of 1.375 for $d_B = 11$ mm. The gap disappears as the thickness ratio approaches to 0.125 corresponding to $d_B = 1$ mm. Also as shown in Fig. 11, the depth of the band gap decreases as the thickness ratio of the layers increases. It is interesting to note that the width and depth of the gap are affected by $\gamma$ but the central frequency is nearly insensitive to the loss factor.

## 4. Conclusion

The effects of loss factor on the characteristics of the zero-$\bar{n}$ gap in a 1D PC consisting of layers with negative and positive indices for different incidence angles and polarizations were studied. At the beginning we showed that the width of the band gap is almost independent of the number of periods for $N \geq 16$. The results show that the width of the band gap increases when the loss factor increases and the gap is omnidirectional and nearly unaffected by the loss factor for TE waves. It was also shown that the zero-$\bar{n}$ gap is very sensitive to the incidence angle and polarization. For TM waves the gap disappears for incidence angles greater than $\approx 50°$. Moreover, the width and the depth of the band gap are very sensitive to the loss factor but the central frequency is not. The effects of the layers thickness ratio and the loss factor on the zero-$\bar{n}$ gap for normal incidence was studied. The results show that the central frequency of the zero-$\bar{n}$ gap decreases as the thickness ratio increases and the depth of the band gap decreases as the thickness ratio of the layers increases. The bandwidth of the zero-$\bar{n}$ gap depends nonlinearly on the thickness ratio and indicates a maximum at 1.375 and disappears at 0.125. The results show that the thickness ratios of the layers play an important role in shifting the central frequency and the shape of the zero-$\bar{n}$ gap. These properties can provide a method for designing new types of optical devices, such as an edge filter in microwave engineering.